\documentclass[twocolumn,aps,pra,nofootinbib,groupedaddress,amsmath,amssymb,longbibliography]{revtex4-1}
\usepackage{graphicx}
\usepackage{dcolumn}
\usepackage{bm}
\usepackage{color}
\usepackage{multirow}
\usepackage[mathscr]{euscript}
\usepackage{subfigure}
\usepackage[colorlinks=true,allcolors=blue]{hyperref}
\usepackage{braket}
\usepackage{lipsum}
\newcommand{\RNum}[1]{\uppercase\expandafter{\romannumeral #1\relax}}

\newcommand{\BE}{\begin{eqnarray}}
\newcommand{\EE}{\end{eqnarray}}
\newcommand{\dagg}{^{\dagger}}
\newcommand{\1}{&=&}

\newcommand{\ee}[1]{\mathrm{e}^{#1}}
\newcommand{\nn}{\nonumber}

\newcommand{\figtop}[4]{
 \begin{figure}[!t]
 \begin{center}
 \scalebox{#3}{\includegraphics{#1}}
 \vspace{-0.1in}
 \caption{\label{#4}#2}
 \end{center}
 \end{figure}
}

\begin{document}
\title{Hyperchaos in a Bose-Hubbard chain with Rydberg-dressed interactions}

\author{Gary McCormack$^{1}$, Rejish Nath$^{2}$ and Weibin Li$^{1}$}

\affiliation{$^{1}$School of Physics and Astronomy, and Centre for the Mathematics and Theoretical Physics of Quantum Non-Equilibrium Systems, University of Nottingham, Nottingham, NG7 2RD, United Kingdom\\ $^2$Department of Physics, Indian Institute of Science Education and Research, Pune 411 008, India}%

\begin{abstract}
We study chaos and hyperchaos of Rydberg-dressed Bose-Einstein condensates (BECs) in a one-dimensional optical lattice. Due to the long-range soft-core interaction between the dressed atoms,  the dynamics of the BECs are described by the extended Bose-Hubbard model. In the mean-field regime, we analyze the dynamical stability of the BEC by focusing on the groundstate and localized state configuration. Lyapunov exponents of the two configurations are calculated by varying the soft-core interaction strength, potential bias and length of the lattice. Both configurations can have multiple positive Lyapunov exponents, exhibiting hyperchaotic dynamics. We show the dependence of the number of the positive Lyapunov exponents and the largest Lyapunov exponent on the length of the optical lattice. The largest Lyapunov exponent is directly proportional to areas of phase space encompassed by the associated Poincar\'e sections. We demonstrate that linear and hysteresis quenches of the lattice potential and the dressed interaction lead to distinct dynamics due to the chaos and hyperchaos. Our work is relevant to current research  on chaos, and collective and emergent nonlinear dynamics of BECs with long-range interactions.
\end{abstract}

\maketitle

\section{Introduction}\label{sec:Introduction}
Over the past two decades, Bose-Einstein condensates (BECs) of ultracold atomic gases have become an ideal system to study both quantum and nonlinear dynamics, due to the high controllability over the two-body interactions~\cite{Smith2013}, trapping potentials~\cite{Pethick2008a} and spatial dimensions~\cite{Fallani2005,Smerzi1997}, along with long coherence times. The emerging nonlinear phenomena depend strongly on the two-body interactions between atoms. In the presence of s-wave interactions, BECs can form dark and bright soliton~\cite{Ma2016,Anderson2001,Dutton2001,Denschlag2000,Burger1999,Cornish2006,Khaykovich2002, Strecker2002} and exhibit Newton's cradle behavior~\cite{Kinoshita2006}, which are paradigmatic examples in nonlinear physics.  In trap array and optical lattice settings, self-trapping of the BEC emerges due to strong repulsive interactions~\cite{Xia2006a,Liu2007,Graefe2006a,Viscondi2011,Li2018d,Chong2005a,Liu2002a,Liu2003a,Albiez2005,Zibold2010}, where the BEC is localized in a single site. This is in contrast to the homogeneous, superfluid state, which form the groundstate of an infinite lattice when the interaction is weak~\cite{Gotlibovych2016,Gaunt2013,Schmidutz2014}. Both the homogeneous and self-trapped states correspond to solutions, i.e.  \textit{fixed points}, of the discrete Gross-Pitaevskii (GP) equation~\cite{Buonsante2008}, which is a nonlinear Schr\"odinger equation that governs the mean-field dynamics. The stability of these fixed points depend on various parameters, such as the s-wave interaction. It has been shown that the self-trapped state in a double-well potential can only be stable when the onsite interaction strength is much stronger than the tunneling strength~\cite{Liu2002a}. Nonetheless, the homogeneous state can be disturbed by the s-wave interaction and external potentials, giving rise to chaotic dynamics~\cite{Hai2008,sinha_chaos_2020}. 
Under strong periodic modulation of the hopping, extended chaotic regions are found in phase space~\cite{Boukobza2010a}.

\figtop{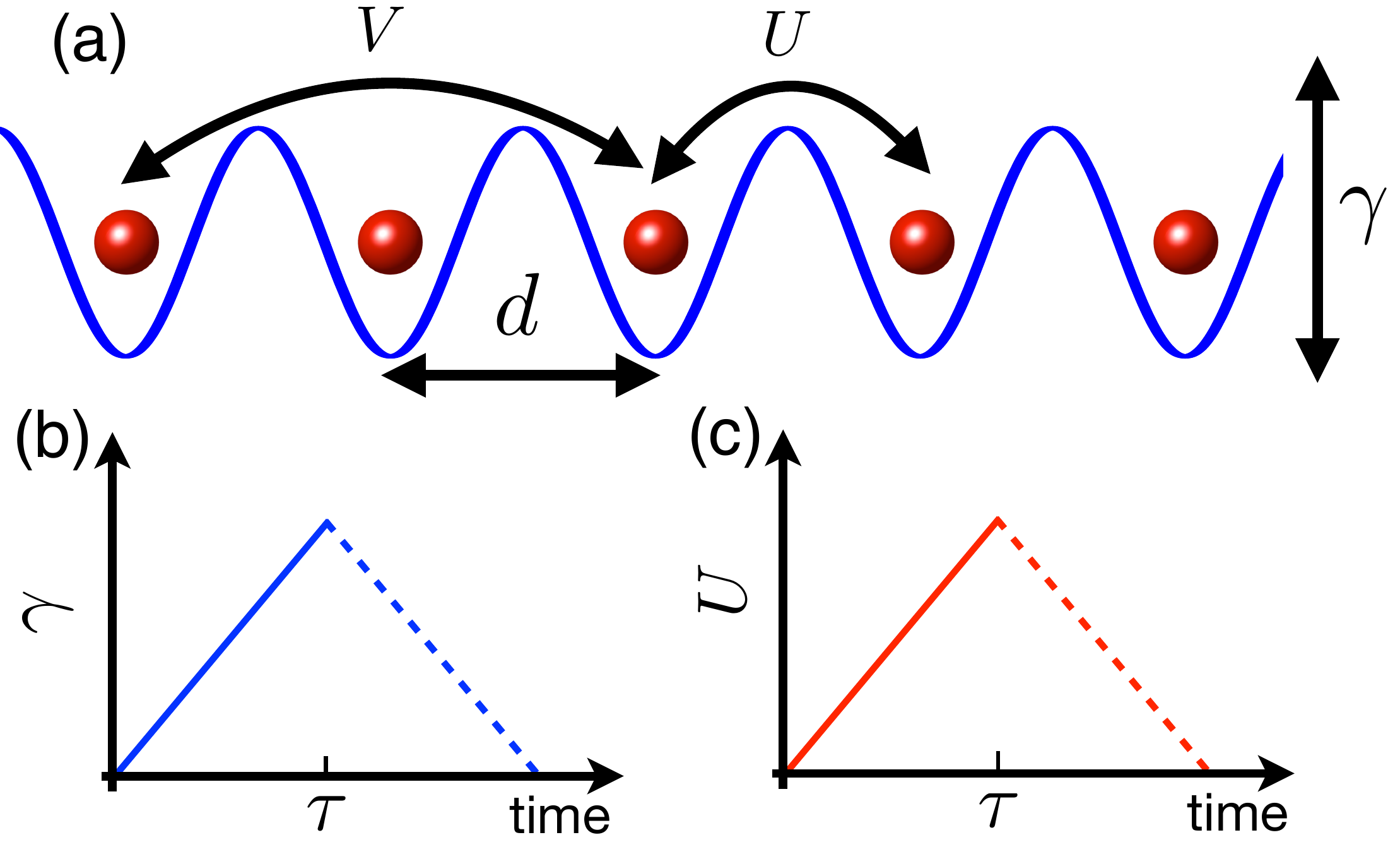}{(Color online) \textbf{The extended Bose-Hubbard chain and quenching schemes.} (a) Nearest-neighbor ($U$) and next-nearest-neighbor ($V$) interactions between atoms in a one dimensional optical lattice (lattice constant $d$). The tilting of the lattice is denoted by the parameter $\gamma$. We consider a linear quench in (b) $\gamma$  and (c) $U$ towards a non-zero value (solid). When $\gamma$ ($U$) returns to the initial value  (both solid and dashed), this is a hysteresis quench. The rate to quench $\gamma$ ($U$) is $\alpha$ ($\beta$) . See text for details of the soft-core interaction and quenching protocols.}{0.35}{fig:intro}
  
On the other hand, long-range interactions play important roles in determining the dynamical stability of BECs. Solitons may occur in BECs in the presence of dipolar interactions~\cite{pedri_two-dimensional_2005-1,tikhonenkov_anisotropic_2008,nath_phonon_2009,cuevas_solitons_2009,young-s_dynamics_2011}. The competition between s-wave and dipolar interactions~\cite{Lahaye2010,Xiong2013,Gallemi2013} leads to bifurcations of the eigenspectra and chaotic dynamics, when confined in harmonic trap~\cite{Koberle2009,Andreev2021_DBEC}. Self-trapping of dipolar BECs in double-well~\cite{xiong_symmetry_2009,abad_dipolar_2011,wang_effects_2011,adhikari_self-trapping_2014} and triple-well potentials~\cite{zhang_dipolar-induced_2012,fortanier_dipolar_2013} have been examined theoretically. Besides the dipolar interaction, one can 
laser couple groundstate atoms to high-lying Rydberg states~\cite{Bouchoule2002,Henkel2010a,Honer2010,Pupillo2010,Johnson2010,Li2012,PhysRevA.93.022709,Hsueh2020}, which induces a long-range \textit{soft-core} interaction between two dressed atoms (with a distance $r$). The soft-core interaction is constant when $r$ is within the soft-core radius $R$, typically of the order of several micrometers~\cite{Henkel2010a}. For $r>R$, the interaction decreases rapidly as $r^{-6}$, shown in Fig.~\ref{fig:intro}(a).
Various theoretical studies on the static and dynamical properties of Rydberg-dressed atoms confined in harmonic traps~\cite{Maucher2011c,Cinti2014,Hsueh2016,McCormack2020} and optical lattices~\cite{Lauer2012,Lan2015,Angelone2016,Chougale2016,Li2018a,Zhou2020,PhysRevA.99.033602} have been conducted in the past decade. 
Rydberg dressed interactions have been experimentally demonstrated in optical tweezers~\cite{Jau2016b}, optical lattices~\cite{Zeiher2016,Zeiher2017,Guardado-Sanchez2020}, and harmonic traps~\cite{Borish2020}. In Ref.~\cite{McCormack2020b} we have shown that self-trapping dynamics of Rydberg-dressed BECs can be controlled in a triple-well potential through mean-field and quantum mechanical analysis.

In this work, we investigate chaotic properties of Rydberg-dressed BECs in a one-dimensional (1D) optical lattice in which the dressed interaction leads to a multi-site density-density interaction. In the semiclassical regime, the nonlinear dynamics of the Bose-Hubbard model is captured by a  discrete, coupled GP equation. Nonlinear eigenenergies, Bogoliubov spectra as well as Lyapunov exponents of the dressed BEC in the lattice are investigated. We then explore dynamical stability of the groundstate and localized state, where dependence of the largest, and total number of positive Lyapunov exponents~\cite{Wolf1985,Andreev2021} on the dressed interaction and system size is explored. We probe the chaotic dynamics by employing both linear and a hysteresis quench of the potential bias and dressed interaction~\cite{Eckel2014,Trenkwalder2016,Burkle2019}. 

The paper is organized as follows. In Sec. \ref{sec:Hamiltonian} the Hamiltonian of the Bose-Hubbard chain is introduced. The corresponding mean-field approximation and GP equations are given. Methods on calculating the eigenenergy, Bogoliubov spectra and Lyapunov exponents are briefly introduced. Quench schemes of the potential bias and nonlinear interaction are explained. 
We explore static (eigenenergies and Bogoliubov spectra) and dynamical properties (Lyapunov exponents) of the groundstate and localized state configurations in  Sec.~\ref{sec:Dynamics_GS}, and Sec.~\ref{sec:Dynamics_Trapped}, respectively. Dynamics driven by both the linear and hysteresis quenching parameters are explored with different initial states.  
In Sec.~\ref{sec:Size} we examine the scaling of the Lyapunov exponents with the system size for the two different configurations. We demonstrate through numerical calculations that areas of the Poincar\'e sections are proportional almost linearly to the largest Lyapunov exponent. 
We conclude our work in Sec. \ref{sec:Conclusion}.

\section{Model and Method}\label{sec:Hamiltonian}
\subsection{Extended Bose-Hubbard model in the semiclassical limit} 
Our setting consists of $N$ bosonic atoms confined in a one-dimensional lattice with lattice constant $d$, as depicted in Fig.~\ref{fig:intro}(a). The Rydberg-dressing induces long-range interactions between atoms at different sites. Taking into account of hopping between nearest-neighbor sites, we obtain an extended Bose-Hubbard Hamiltonian of $L$ sites~\cite{Li2012} ($\hbar=1$)
\BE
\label{Ham:bh}
\hat{H} &=& -J\sum_{\langle i,j\rangle}^{L}\hat{a}_i\dagg\hat{a}_j +\sum_j^{L}\Gamma_j\hat{n}_j +\frac{1}{2}\hat{H}_{\text{int}},
\label{eq:eBHM}
\EE
where $\hat{a}_j (\hat{a}_j^{\dagger})$ is the bosonic annihilation (creation) operator at site $j$. The tunneling strength $J$ acts only on nearest-neighbor sites, denoted by $\langle\cdot\rangle$ in the summation. Here, $\hat{n}_j=\hat{a}_j\dagg\hat{a}_j$ is the number operator, while $\Gamma_j$ is the local tilting potential. The titling is given by $\Gamma_j=-\gamma\left(j-1-\lfloor L/2 \rfloor\right)$, where $\lfloor\cdot\rfloor$ and $\gamma$ are the floor function and level bias between neighboring sites, respectively.
The onsite and long-range interactions are described by $\hat{H}_{\text{int}}= g\sum_{j}^{L}\hat{n}_j\left(\hat{n}_j - 1\right)+ \sum_{i,j}^{L}\Lambda_{i,j}\hat{n}_i\hat{n}_j$. The onsite interaction $g=4\pi a_s/m$~\cite{Pethick2008a} depends on the s-wave scattering length $a_s$ and mass $m$, where the former can be adjusted by Feshbach resonances~\cite{Pethick2008a}. The soft-core shaped long-range interaction is given by
$\Lambda_{i,j}=C_6/[|i-j|^{6}d^{6} + R^{6}]$ with $C_6$ being the dispersion coefficient [Fig.~\ref{fig:intro}(a)]. Both the soft-core radius $R$ and $C_6$ can be tuned by laser parameters~\cite{Henkel2010a}. In this work, we will restrict to the onsite, nearest-neighbor ($\Lambda_{j,j\pm1}$) and next-nearest-neighbor  ($\Lambda_{j,j\pm2}$) interactions only,  where $R\sim d$. This approximation is valid as the soft-core interaction decays rapidly when the separation between sites is larger than the soft-core radius.

In the semiclassical limit $N\gg 1$, we employ the mean-field approximation where the bosonic operator is described by a classical field $\psi_{j}$, i.e. $\hat{a}_j\approx\psi_j\sqrt{N}$, and  $\hat{a}_j^{\dagger}\approx\psi_j^{*}\sqrt{N}$, with the normalization condition $\sum_{j}|\psi_j|^2=1$. This yields the semiclassical Hamiltonian $\mathcal{H}\approx\hat{H}/N$, 
\BE
\label{Ham:mf}
\mathcal{H}&=&\sum_j^L\Gamma_j |\psi_{j}|^2-J\sum^L_{j}\left(\psi_{j+1}^*\psi_{j}+\psi_{j-1}^*\psi_{j}\right)\nn \\
&&+\frac{N}{2}\sum_{i,j}^{L}|\psi_j|^2\left[g\left(|\psi_j|^2 - 1\right)+\Lambda_{i,j}|\psi_i|^2\right].
\EE
The dynamics of the classical field $\psi_j$ is obtained via the canonical equation $id\psi_{j}/dt=\partial \mathcal{H}/\partial \psi_{j}^*$, yielding the coupled GP equations
\BE
\label{EoM}
i \dot \psi_j &=&- J\left(\psi_{j+1}+\psi_{j-1}\right)+\big[\Gamma_j + W|\psi_j|^2 +\\ 
	&& U\left(|\psi_{j+1}|^2+|\psi_{j-1}|^2\big)+V\big(|\psi_{j+2}|^2+|\psi_{j-2}|^2 \right)\big]\psi_j, \nonumber 
\EE
where we have defined $W=N(\Lambda_{j,j} + g)$, $U=N\Lambda_{j,j\pm1}$, and $V=N\Lambda_{j,j\pm2}$, to be the onsite, nearest-neighbor and next-nearest-neighbor  interaction strength. The onsite interaction $W$ takes into account  contributions from both the s-wave and soft-core interaction. We will assume a vanishing onsite interaction, i.e. $W=0$ which allows us to focus on effects induced by the long-range interaction part. To be concrete, we will fix the nearest-neighbor and next-nearest-neighbor interaction to be $U=2V$ in the following discussion.  Time and energy will be scaled with respect to $1/J$ and $J$ in what follows.

It is convenient to examine the real ($\mathcal{R}_j=\rm{Re}[\psi_j]$) and imaginary components ($\mathcal{I}_j=\rm{Im}[\psi_j]$) of $\psi_j$, 
\BE
\dot{\mathcal{R}}_j &=& -J\left(\mathcal{I}_{j+1}+\mathcal{I}_{j-1}\right) +\big[\Gamma_j + W|\psi_j|^2 +\\ 
&& U\left(|\psi_{j+1}|^2+|\psi_{j-1}|^2\big)+V\big(|\psi_{j+2}|^2+|\psi_{j-2}|^2 \right)\big]\mathcal{I}_j,\nonumber\label{eqn:imag}\\
\dot {\mathcal{I}}_j &=&  J\left( \mathcal{R}_{j+1}+\mathcal{R}_{j-1}\right) -\big[\Gamma_j + W|\psi_j|^2 +\\ 
&& U\left(|\psi_{j+1}|^2+|\psi_{j-1}|^2\big)+V\big(|\psi_{j+2}|^2+|\psi_{j-2}|^2 \right)\big]\mathcal{R}_j,\nonumber\label{eqn:imag2}
\EE
with $|\psi_j|^2=\mathcal{R}_j^2+\mathcal{I}_j^2$. Both $\mathcal{R}_j$ and $\mathcal{I}_j$ are real valued functions of time. We will calculate Lyapunov exponents and the Poincar\'e sections based on these real functions. Note that  $\mathcal{R}_j$ and $\mathcal{I}_j$ represent mean values of the quadrature of the operator $\hat{a}_j$.  The quadrature fulfills the commutation relation similar to the position and momentum operator~\cite{scully_quantum_1997}. Hence the mean values of the quadrature allow us to obtain useful information on the dynamics of the system in phase space.  For small systems, $L=2$ or $3$, one can also describe the classical field with the canonical phase and particle number decomposition~\cite{Liu2007,castro2021quantumclassical}. 

\subsection{Nonlinear eigenenergies and Bogoliubov spectra}\label{sec:Energetic_rymma}
Though the Hamiltonian~(\ref{Ham:mf}) is Hermitian, the density-dependent nonlinearity prevents us from calculating the eigenenergy through conventional diagonalization. To overcome this, a shooting method will be employed to numerically evaluate the eigenstate $\bar{\Psi}_j=[\bar{\psi}_1,\bar{\psi}_2,\cdots,\bar{\psi}_L]$ and corresponding eigenenergy $\varepsilon_j$ self-consistently~\cite{McCormack2020b}. A trial solution is seeded into the semiclassical Hamiltonian. It is then diagonalized,  leading to a new eigenstate and eigenenergy. This process is iterated until the resulting eigenstate and eigenenergy is obtained self-consistently.

For interacting systems, one can analyze the Bogoliubov spectra $\epsilon_B$ to understand the stability of the eigenstate. This is achieved by linearizing around a given state $\bar{\Psi}$ (e.g., a fixed point of the semiclassical system), where each component is given by $
\psi_j = \bar{\psi}_j +u_j \ee{-i \epsilon_{B}t} -v_j^*\ee{i\epsilon_{B} t},\label{eq:fluc}$ with $u_j$ and $v_j$ being the probability amplitudes of the Bogoliubov quasiparticles~\cite{Pethick2008a}.  The dynamics of $u_j$ and $v_j$ are described by the Bogoliubov equations~\cite{Dey2018,Dey2019},
\BE
\left(
\begin{matrix}
	\mathcal{L} & \mathcal{N}\\ 
	-\mathcal{N} & -\mathcal{L}
\end{matrix}
\right) 
\left(
\begin{matrix}
	\mathbf{u}\\ \mathbf{v}
\end{matrix}
\right)
\1 \epsilon_{B}
\left(
\begin{matrix}
	\mathbf{u}\\ \mathbf{v}
\end{matrix}
\right)
\EE
where $\mathcal{L}=\tilde{H}_0+2 U \mathcal{P}-\mu$, and $\mathcal{N} = -U\mathcal{P}$. $\tilde{H}_0$ and $\mathcal{P}$  are $L\times L$ block matrices. From Eq.~(\ref{EoM}), we obtain the matrix elements  $\langle \psi_j|\tilde{H}_0|\psi_j \rangle = \Gamma_j$, $\langle \psi_j|\tilde{H}_0|\psi_{j\pm 1} \rangle  =  -J$,
 $\langle\psi_j|\mathcal{P}|\psi_j \rangle = |\bar{\psi}_{j+1}|^2+|\bar{\psi}_{j-1}|^2 + (  |\bar{\psi}_{j+2}|^2+|\bar{\psi}_{j-2}|^2)/2$, $\langle\psi_j|\mathcal{P}|\psi_{j\pm1} \rangle = 2\bar{\psi}_{j\pm1}\bar\psi_{j}$ and $\langle \psi_j|\mathcal{P}|\psi_{j\pm2} \rangle = \bar{\psi}_{j\pm2}\bar\psi_{j}$, while other matrix elements are zero. If the Bogoliubov spectra are complex numbers, the state is then \textit{dynamically unstable}, as Bogoliubov quasiparticles grow (decay) exponentially with time, whose rate is determined by the imaginary part of the spectra.  

\subsection{Poincar\'e sections and Lyapunov exponents}\label{sec:Lyap_Gamma}
The emergence of chaos in the dynamics can be characterized by the Poincar\'e sections and Lyapunov exponents. For $L$ sites, the possible trajectories are the complete set of $\{\mathcal{R}_1,\cdots,\mathcal{R}_L,\mathcal{I}_1,\cdots,\mathcal{I}_L \}$. Due to the normalization condition, we need to solve a $2L-1$ dimensional system to obtain the dynamics. It is difficult to comprehend the stability of the trajectories in such a high dimensional phase space. Instead, we project the dynamics to a two dimensional (2D) Poincar\'e section to identify the dynamical properties.  To calculate the 2D Poincar\'e section, we record trajectories of selected variables ($\mathcal{R}_j$, $\mathcal{I}_j$) as they cut through the $\mathcal{U}_k$-plane ($j\neq k$), provided that $\dot{\mathcal{U}}_k>0$. These intersecting points form the 2D Poincar\'e section. To be specific, we will evaluate the Poincar\'e section of variable ($\mathcal{R}_2$, $\mathcal{I}_2$) on the $\mathcal{U}_1$ plane. 

The strength of chaos can be measured by the Lyapunov exponents associated with the equations of motion~\cite{Wolf1985,Andreev2021}. The Lyapunov exponents give the rate of separation between trajectories for a given initial state. As the Lyapunov exponents depend on the initial state, we will consider both the groundstate and a localized state initially. In a localized state, nearly all the condensate sits in a single site, which can be stable (i.e. the self-trapping state) when the nonlinear interaction is strong. In this work, the Lyapunov exponents $\lambda_j$ ($j=1,\cdots,2L$) are calculated via DynamicalSystems.jl, a fast and reliable Julia library to determine the dynamics of nonlinear systems~\cite{Datseris2018}. We have checked that it gives consistent data with the method in Ref.~\cite{Wolf1985}. 
When there exists at least one positive Lyapunov exponent the trajectories will separate exponentially, leading to chaotic dynamics. The dynamics is hyperchaotic when there are more than two positive Lyapunov exponents~\cite{Andreev2021}. 

\begin{figure}
\includegraphics[width=\linewidth]{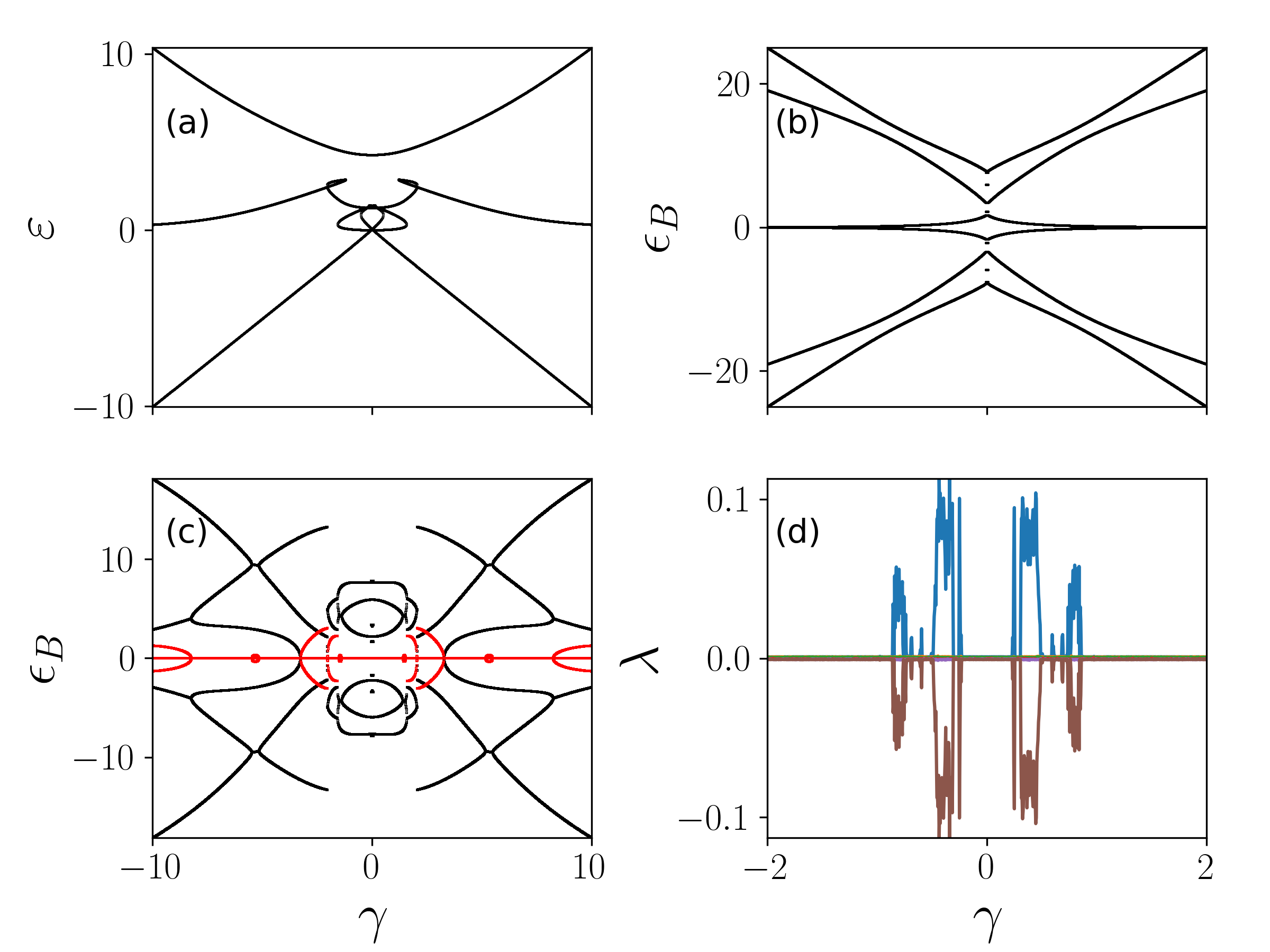}
\caption{\textbf{Eigenenergies, Bogoliubov spectra and Lyapunov exponents when varying the tilt $\gamma$.} We show (a) the nonlinear eigenenergy,  Bogoliubov spectra of (b) the groundstate and (c) the first excited state , and (d) the Lyapunov exponents  of the  groundstate. The nonlinearity dominates when $|\gamma|$ is small, leading to loops in the eigenenergy. The Bogoliubov spectra are all real when the system is in the groundstate (b). The Bogoliubov spectra have complex components (red region) when the system is in the first-excited state. Positive Lyapunov exponents indicate the system exhibits chaos dynamically, which appear mostly in the loop region of the eigenenergy. Parameters are $L=3$ and $U=2V=5$.}
\label{fig:groundstate}
\end{figure}
\subsection{Quenching schemes}
In Sections. \ref{sec:Dynamics_GS} and \ref{sec:Dynamics_Trapped} we will explore the dynamics of the system with time-dependent parameters via the following quenching schemes. 

\textbf{\textit{Scheme $\mathbf{I}$}}: First we consider a linear quench of the potential bias~\cite{McCormack2020b}. The bias between two neighboring sites is given by the function 
\BE
\gamma_{L} = \gamma_i + \alpha t, \label{eqn:GammaL}
\EE
where $\gamma_i$ and $\alpha$ are the initial value and quench rate, respectively. With $\gamma_i<0$, the quench takes place from $t=0$ to $t=2\gamma_f/\alpha$ with $\gamma_f = -\gamma_i$, depicted by the solid curve in Fig.~\ref{fig:intro}(b).

\textbf{\textit{Scheme $\mathbf{II}$ }}:
Alternatively, we consider a hysteresis quench~\cite{Eckel2014,Trenkwalder2016,Burkle2019} where the system begins at $\gamma_i$ and then evolves to $\gamma_f$. At time $\tau=\gamma_f/\alpha$, the potential bias is quenched back towards $\gamma_i$. The function describing this scheme is 
\BE
\gamma_{H} = \gamma_f +(\gamma_i-\gamma_f) \frac{|\tau-t|}{\tau}. \label{eqn:GammaH}
\EE
The corresponding scheme is shown by the solid and dashed curve in Fig.~\ref{fig:intro}(b). 

\textbf{\textit{Scheme $\mathbf{III}$ }}:
In addition to quenching the level bias we also change the two-body interaction strength through  a linear ramp,
\BE
U_{L} = U_i +\beta t \label{eqn:UL}
\EE
where $U_i$ is the initial interaction strength and $\beta$ is the quench rate. This is shown by the solid curve in Fig.~\ref{fig:intro}(c). Note that the next-nearest-neighbor interaction $V$ depends on time as well due to the relation $U=2V$. 

\textbf{\textit{{Scheme $\mathbf{IV}$ }}}:
The hysteresis counterpart of the interaction quench is given by
\BE
U_{H}=U_f+(U_i-U_f) \frac{|\tau'-t|}{\tau}, \label{eqn:UH}
\EE
where $U_f$ is the final interaction strength, with $\tau'=U_f/\beta$. 
\begin{figure}
	\includegraphics[width=\linewidth]{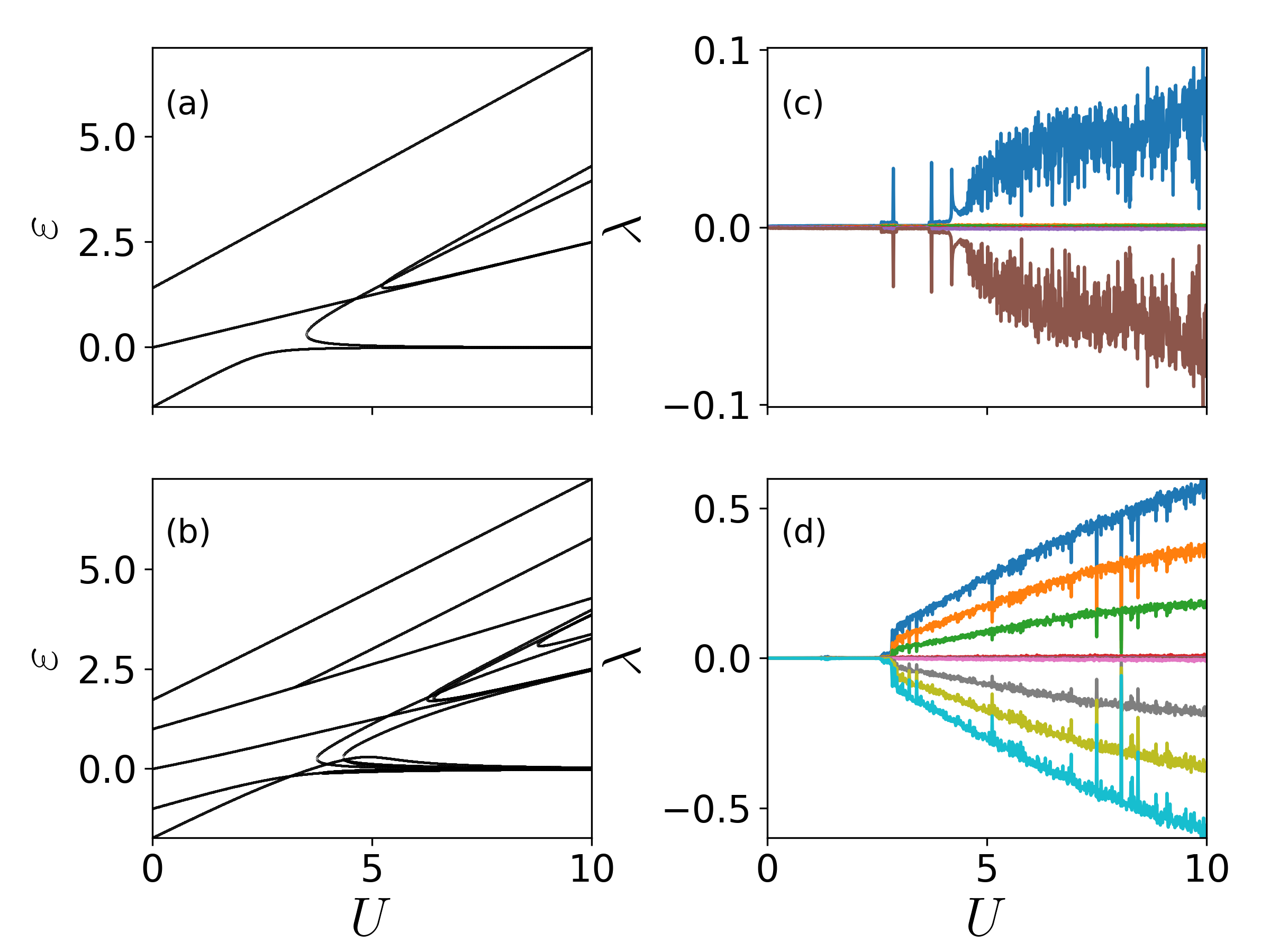}
	\caption{\textbf{Eigenenergies and Lyapunov exponents as a function of $U$.} We show eigenenergy for (a) $L=3$ and (b) $L=5$ when the trap is balanced ($\gamma=0$). Level crossings are found when the interaction is strong. Starting from the groundstate, we calculate Lyapunov exponents for (c) $L=3$ and (d) $L=5$. For a given $U$, Lyapunov exponents of same value but opposite signs appear in pairs. }
	\label{fig:spectra}
\end{figure}
\section{Stability of the groundstate} \label{sec:Dynamics_GS}
\subsection{Eigenenergies, Bogoliubov spectra and Lyapunov exponents}
Without the nonlinearity, the number of eigenenergies $N_{\epsilon}$ is identical to $L$, the dimension of the semiclassical system. The number of eigenenergies can be larger than $L$ when the interaction is strong. As an example, eigenenergies for $L=3$ as a function of the bias $\gamma$ are shown  in Fig.~\ref{fig:groundstate}(a). We find $N_{\epsilon}>L$ when $|\gamma|\lesssim U$, where the nonlinearity dominates. Loops and crossings appear in the eigenenergies, except the highest energy level. 

For a given state of the nonlinear system, one obtains $2L$ Bogoliubov spectra, whose values depend on the specific eigenstate and nonlinear interaction strength. The Bogoliubov modes are stable for all $\gamma$ when the system is in the groundstate, i.e., the Bogoliubov spectra $\epsilon_B$ are real, as shown in Fig.~\ref{fig:groundstate}(b). This is in contrast to excited eigenstates, whose Bogoliubov spectra have imaginary components. As an example, the Bogoliubov spectra of the first excited state is shown in Fig. \ref{fig:groundstate}(c). The corresponding Bogoliubov mode will decay (grow) exponentially, when the imaginary part is negative (positive). 
\begin{figure}
	\centering
	\includegraphics[width=\linewidth]{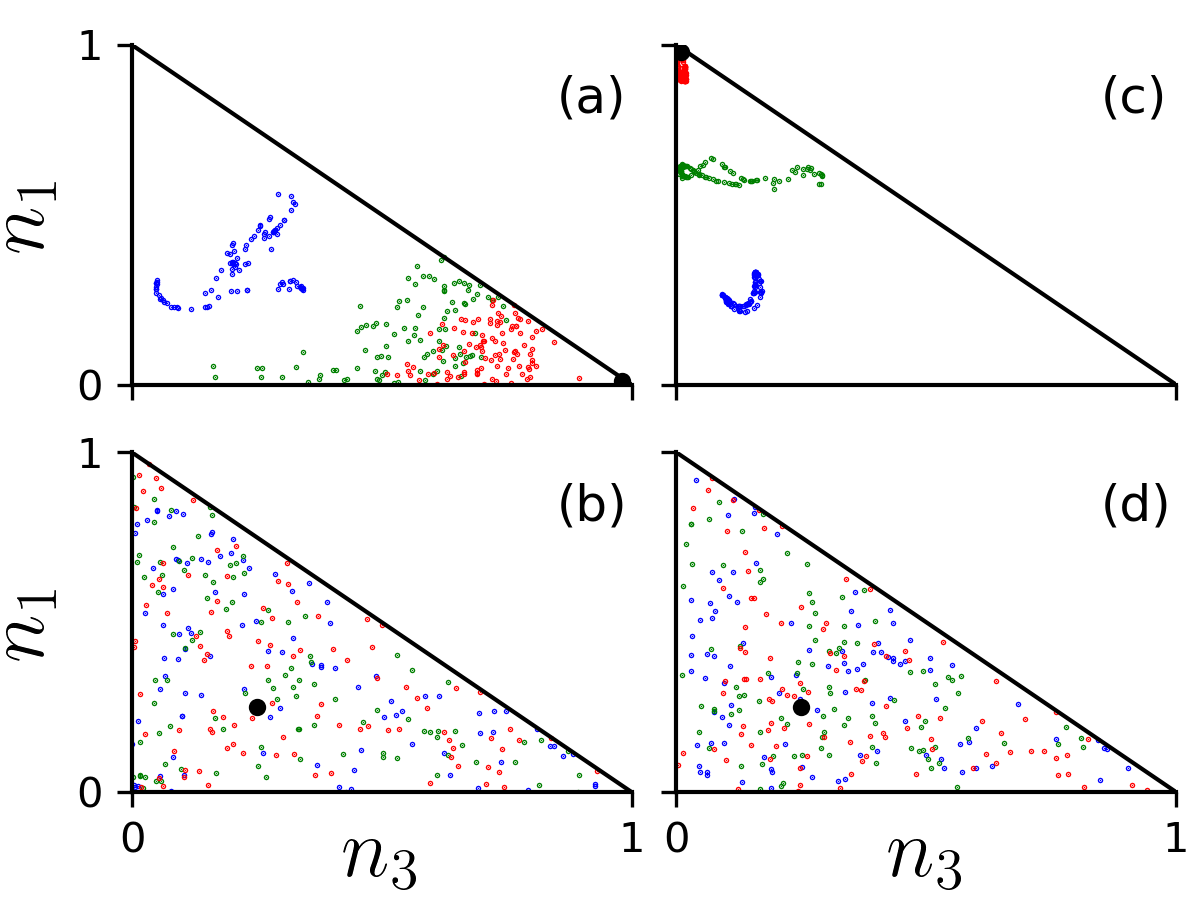}
	\caption{(color online) \textbf{Final population distribution of the groundstate}. The population by quenching $\gamma$ with  (a) scheme $\mathbf{I}$  and (b) scheme $\mathbf{II}$ is shown for $L=3$.  In the numerical simulation, $\gamma_i=-\gamma_f=-10$ and the interaction strength is $U=5$. The interaction $U$ is quenched with (c) scheme $\mathbf{III}$  and  (d) scheme $\mathbf{IV}$where $U_i=0$, $U_f=10$ and $\gamma=0$ , respectively. In all the figures, the quench rates ($\alpha$ or $\beta$) are $1$ (blue), $0.1$ (green), and $0.01$ (red). The total number of trajectories is $M=100$. The target final state is shown as the large black circle. }\label{fig:dynamics_gs}
\end{figure}

The chaotic dynamics of the system is characterized by positive Lyapunov exponents. In Fig.~\ref{fig:groundstate}(d) Lyapunov exponents are shown for the groundstate of the system. When increasing $\gamma$,  negative and positive Lyapunov exponents  are found in regions where the eigenenergies show loops. The negative and positive Lyapunov exponents appear in pairs with the same absolute values, as our system is conservative. In this example, one positive Lyapunov exponent can be found when $|\gamma|<1$, indicating the presence of chaos. This means that small fluctuations on the groundstate could gain exponential growth, and hence drives the system away from the groundstate.  

To further understand roles played by the nonlinearity, we calculate eigenenergies as a function of the interaction strength $U$ shown in Fig. \ref{fig:spectra}(a) and (b), for $L=3$ and $L=5$, respectively. It can be seen that new branches are generated when the nonlinear interaction $U$ is large enough.  Lyapunov exponents of the groundstate of the nonlinear system are shown in Figs. \ref{fig:spectra}(c) and \ref{fig:spectra}(d). Positive Lyapunov exponents are found in the strongly interacting region, whose values increase with increasing $U$. Larger Lyapunov exponents mean that the exponential growth of the instability can be even faster. Importantly, the number of Lyapunov exponents now depends on $L$. For $L=3$, one obtains single positive Lyapunov exponent when $U\gtrsim4$.  
When $L=5$, there are  3 positive Lyapunov exponents. This indicates that the system enters the so-called  \textit{hyperchaos}  regime~\cite{Baier1990,Baier1995,Kapitaniak1995}, where more than one positive Lyapunov exponents can be found in the dynamics.  In the two examples, we obtain maximally $L-2$ positive Lyapunov exponents, as the energy and particle number is conserved in the Bose-Hubbard chain.
\subsection{Quench dynamics}\label{sec:Hyst_Gamma}
In the linear regime, dynamics of the system will follow the eigenstate adiabatically when slowly quenching the tilt potential. However the dynamics may deviate from the adiabatic eigenstate in the nonlinear regime, especially when positive Lyapunov exponents are found. This will be illustrated through quenching the tilt potential and interaction strength given by Eqs. (\ref{eqn:GammaL})-(\ref{eqn:UH}). To trigger the instability in the dynamics, we consider a thermal mixed state $\bar{\Psi}_j'=[\bar \psi_1 \ee{i\theta_1},\bar \psi_2\ee{i\theta_2},\cdots,\bar \psi_L\ee{i\theta_L}]$ around a given state $\bar{\Psi}_j$ (the groundstate), where $\theta_j$ is a random phase distributed uniformly between $0$ and $2\pi$ \cite{Burkle2019}. In numerical simulations, we typically consider an ensemble of $M=100$ realizations with a given set of parameters. 

We first examine a linear quench of the bias $\gamma$ when the system is prepared in the groundstate at $\gamma_i=-10$ and $L=3$. The majority of the condensate is located on the first (leftmost) site [$n_1(0)=|\psi_1|^2\approx 1$] initially [Fig. \ref{fig:dynamics_gs}(a)]. In the adiabatic limit and without the nonlinear interaction, the condensate will move to the third well, $n_3(\tau)\approx1$, after the quench~\cite{McCormack2020b}. The population at this adiabatic limit is shown with a black dot in each panel. 
The quench dynamics however depend on the finite quench rate and the interaction strength. When the interaction is weak the condensate can be in any of the three sites, since the tunneling strength between neighboring sites plays the dominant role.  The distribution of the final population is affected by the noise on the initial state and also depends on the final time in the simulations. Increasing $U$, the population is distributed into a larger region of phase space, i.e., it occupies a larger areas in the $n_3$-$n_1$ plane. By fixing the interaction $U$, our numerical simulation shows that the smaller $\alpha$ is, the closer the population distribution is to the adiabatic limit. 

For the hysteresis quench given by Eq. (\ref{eqn:GammaH}), we see that even for $U=5$ (meaning the eigenstate exhibits complicated level crossings) the density mostly returns to their initial state, at least when $\alpha \ll 1$ [Fig. \ref{fig:dynamics_gs}(b)]. Here the hysteresis quench has allowed for a large level of  \textit{reversibility} in the dynamics~\cite{Burkle2019}, as the chaotic regions have not been triggered. Increasing the quench rate $\alpha$, the population distributions cluster around much smaller regions in phase space, than the one shown in panel (a).

In Fig. \ref{fig:dynamics_gs}(c) we quench the interaction according to Eq. (\ref{eqn:UL}). The initial states depend on the value of $\gamma$. For example the groundstate is $\bar{{\Psi}}=[0.5,1/\sqrt{2},0.5]$ for $\gamma=0$. The final states are highly dependent on the initial conditions, due to the chaos in the dynamics [see the crossing energy levels in Fig. \ref{fig:spectra} (a) and Lyapunov exponent in Fig.~\ref{fig:spectra}(c)].  We have verified that by increasing $\gamma$ the associated randomness with the final states decreases, as the number of crossings in the eigenenergy will decrease. 

In case of the hysteresis quench of $U$, we find that the results [Fig.~\ref{fig:dynamics_gs}(d)] are similar to the linear quench. When looking at $\gamma=0$, the final states do not return to the initial value. As shown in Fig.~\ref{fig:spectra}(c), the Lyapunov exponent of the groundstate  becomes positive when $U\gtrsim4$, which causes the final state more random, i.e. a broader distribution of the densities. As the tilt $\gamma$ increases, we have verified that chaos is gradually suppressed, as the population localizes in the trap corresponds to the lowest energy state. In order to trigger chaotic dynamics in the tilted case, stronger interactions are needed in general. 
\begin{figure}
	\centering
	\includegraphics[width=\linewidth]{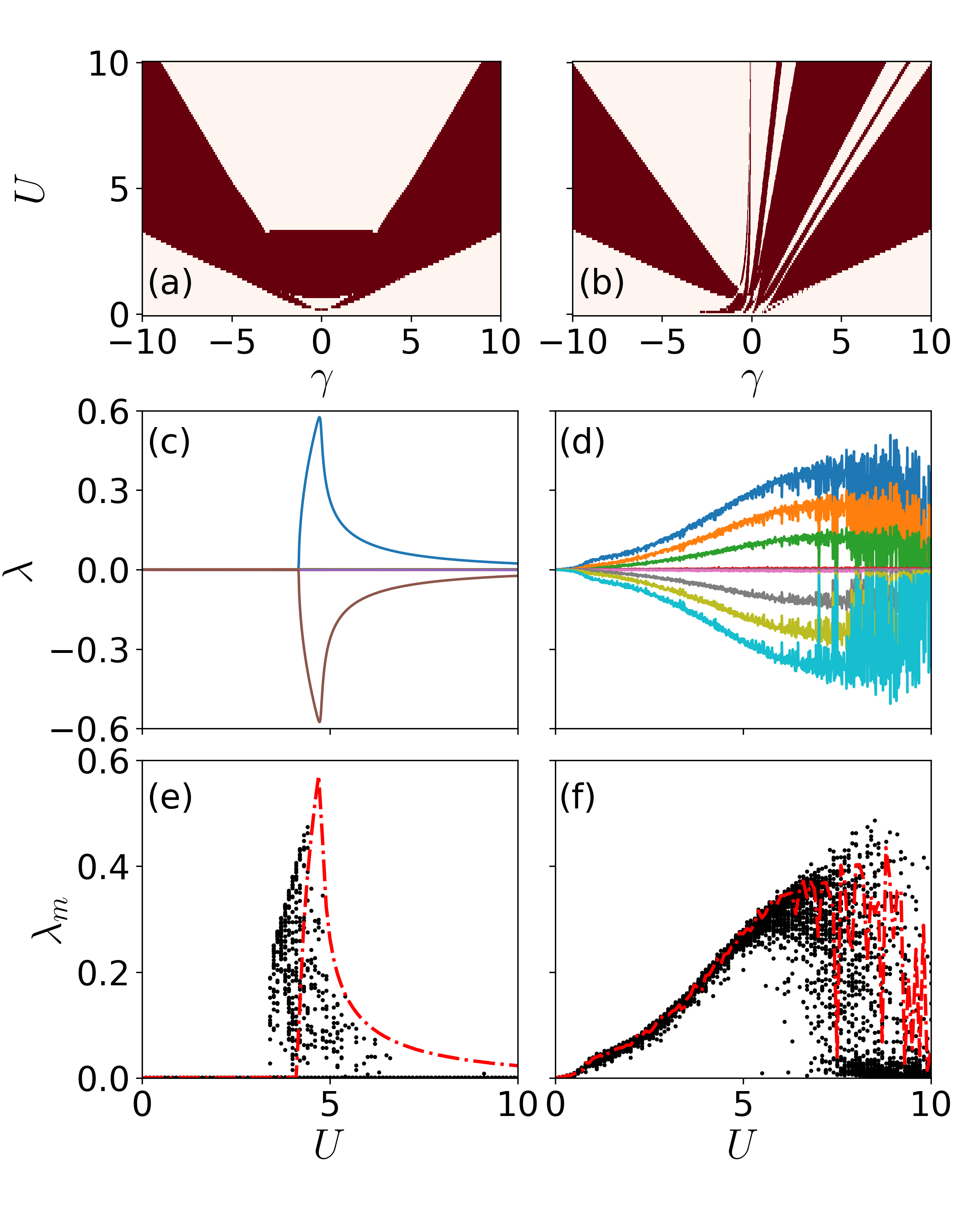} 
	\caption{(color online) \textbf{Bogoliubov spectra and Lyapunov exponents of the localized state.} 
		Dynamically unstable regions (dark red) for (a) $L=3$ and (b) L=5 are shown as a function of $U$ and $\gamma$. Panels (c) and (d) give the Lyapunov exponents as a function of $U$. Random perturbation to the initial state are examined for (e) $L=3$ and (f) $L=5$. The red lines show the maximal Lyapunov exponents in (c) and (d), correspondingly. Here $\gamma=0$ in panels (c)-(d).}
	\label{fig:stability2}
\end{figure}

\begin{figure}[t!]
	\centering
	\includegraphics[width=0.94\linewidth]{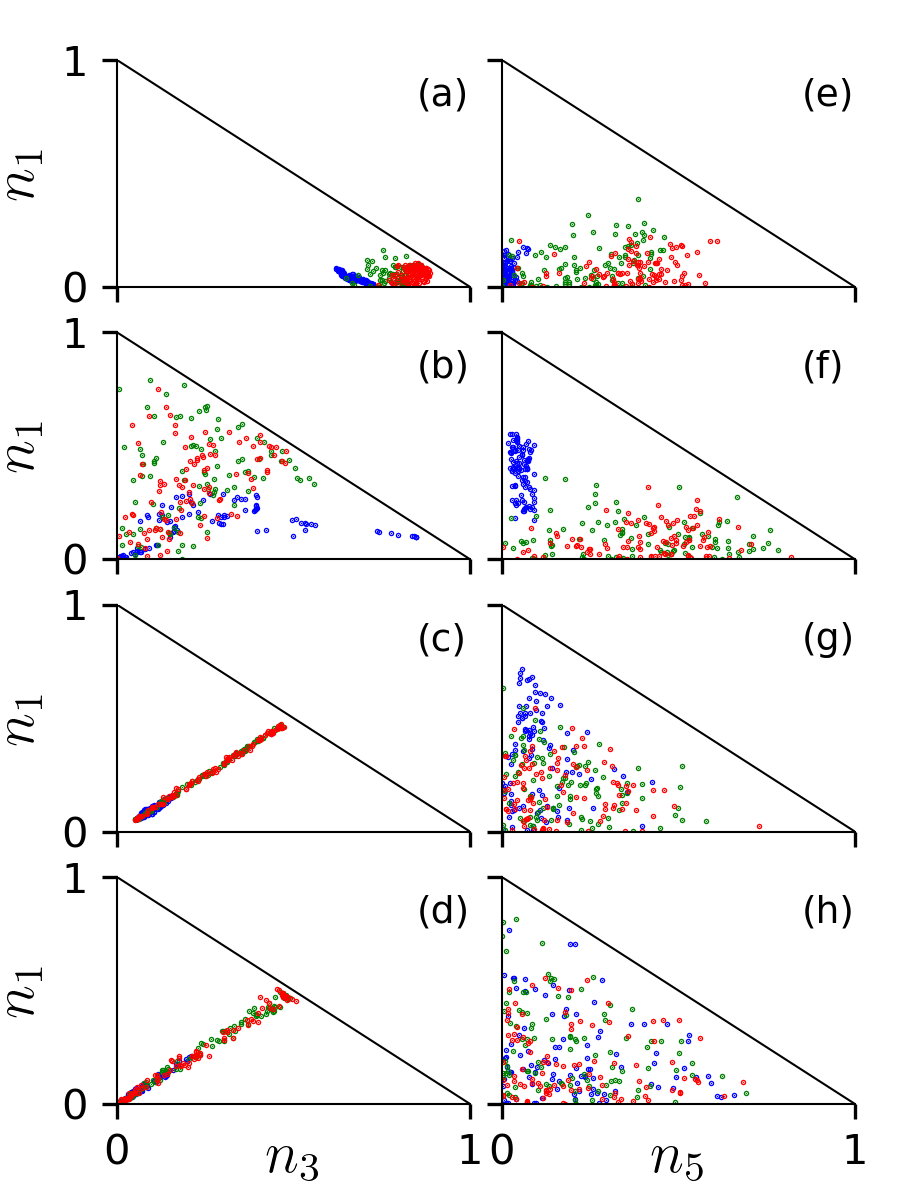}
	\caption{(color online) \textbf{Final population distribution of the localized state}. The first and second row show the linear and hysteresis quench of $\gamma$. Here $U=5$, $\gamma_i=-\gamma_f=10$. The third and fourth  row show the linear and hysteresis quench of $U$ with $U_i=0$, $U_f=10$ and $\gamma=0$.  In (a)-(d) we consider three sites and the initial thermal state is ${\bar{\Psi}}=[0.1e^{i\phi_1},\sqrt{0.98}e^{i\phi_2},0.1e^{\phi_3}]$  with $\phi_j$ ($j=1,\,2,\, 3$) are random number in $[0,2\pi]$.  In (e)-(h) $L=5$ and the initial state is ${\bar{\Psi}}=[\sqrt{0.005}e^{i\phi_1},\sqrt{0.98}e^{i\phi_2},\sqrt{0.005}e^{i\phi_3},\sqrt{0.005}e^{i\phi_4},\sqrt{0.005}^{i\phi_5}]$ with $\phi_j$ ($j=1,\cdots,5$) being randomly distributed in $[0,2\pi]$. The small fraction in sites other than the localized state is used to trigger the hopping dynamics.  Other parameters are same with the one in Fig.~\ref{fig:dynamics_gs}.}
	\label{fig:dynamics_as}
\end{figure}

\section{Stability of the localized state}\label{sec:Dynamics_Trapped}
\subsection{Bogoliubov spectra and Lyapunov exponents}
In this section, we will explore stability of a situation where the condensate is trapped in a single site. When localized at one end of the lattice, it corresponds to the groundstate if the lattice potential is strongly tilted $|\gamma|\gg 1$. We will examine dynamics of localized states even in the balanced case ($\gamma=0$), partially motivated by the fact that the self-trapped state can be stabilized by strong nonlinear interactions. We will show that dynamical instabilities of localized states will depend strongly on the long-range interaction. To be concrete, we will consider a scenario where the condensate is confined in the second trap from the left of the lattice, i.e.  $\bar{\Psi} = [0,1,\cdots,0]$. In the numerical simulations of the dynamics, uniform density fluctuations are applied to the lattice to trigger the hopping dynamics. This modifies the initial state to be
$\bar{\Psi} = [\sqrt{\varepsilon/L}e^{\phi_1},\sqrt{1-\varepsilon}e^{\phi_2},\cdots,\sqrt{\varepsilon/L}e^{\phi_L}]$ with $\epsilon\ll 1$ and $\phi_j$ to be a random phase. This choice furthermore  insures that the energy of different initial states are almost identical. 

In Figs. \ref{fig:stability2}(a) and (b), dynamical unstable regions in the Bogoliubov spectra for $L=3$ and $L=5$ are shown (highlighted with dark red color). In the unstable region, $\epsilon_B$ develops imaginary components,  which depend on $U$, $\gamma$ and $L$. In case of $L=3$, the condensate is localized in the middle site initially, meaning the Bogoliubov spectra are symmetric with respect to $\gamma$. Fig.~\ref{fig:stability2}(a) shows that the system is dynamically unstable when $U$ is small, in particular when the lattice is balanced ($|\gamma|$ is small). This is not surprising, as the localized state is not the groundstate, nor the system supports the self-trapped state. By increasing the interaction strength, we note that the localized state returns to a stable configuration when $|\gamma|$ is small. This means that the localized state becomes a stable, self-trapped state~\cite{McCormack2020b}. When $L=5$, the dynamical stability now depends heavily on tilt $\gamma$. When $\gamma>0$ there is a much broader range of unstable regions. This feature is largely due to that the nonsymmetric initial state has higher energies. Therefore we expect to see qualitatively different dynamics from the various quenching schemes.

The Lyapunov exponents exhibit sensitive dependence on the system size. As shown in Fig.~\ref{fig:stability2}(c) the Lyapunov exponents for $L=3$ show an unusually symmetric shape when $U\gtrsim4$. The exponents are a smooth function of $U$, and reaches maximal value around $U=5$. Further increasing $U$, the positive Lyapunov exponents decrease. This indicates that the localized configuration could exhibit chaotic dynamics for large $U$. 
For $L=5$ we notice that positive Lyapunov exponents can be found when $U$ is relatively small. A key difference is that there are multiple positive Lyapunov exponents  [Fig. \ref{fig:stability2}(d)], where the nonlinear dynamics enters the hyperchaotic regime. 

To understand the maximal Lyapunov exponents, we slightly alter the initial state so that we have $\bar\Psi=\left[\varepsilon,\sqrt{1-2\varepsilon^2},\varepsilon,\cdots,0 \right]$, where $\varepsilon$ is a small perturbation to the wavefunction of the traps on either side of the localized site, with $0<\varepsilon<0.01$. 
In Fig. \ref{fig:stability2}(e) and (f) [corresponding to $L=3$ and $5$] the largest Lyapunov exponent $\lambda_m$ (red) and Lyapunov exponents obtained with modified initial states (black) are shown (only the positive branch).  It shows that a minor change to the initial state will change Lyapunov exponents significantly. However $\lambda_m$ gives an approximate upper bound for all the Lyapunov exponents.  

\subsection{Quench dynamics}
For $U=5$ and $L=3$, a linear quench [Fig.~\ref{fig:dynamics_as}(a)] from $\gamma_i=-10$ to $\gamma_f=10$ shows strong self-trapping behavior in the rightmost potential. Ideally we would expect that by performing a hysteresis quench back towards $\gamma_i$, the population would localize in the leftmost site again. However from Fig.~\ref{fig:dynamics_as}(b) we see that the final state is rather chaotic. Due to the dynamical instability and chaos near $|\gamma|<1$, the final state deviates from the initial state. In panels (c) and (d) we quench according to Eqs.~(\ref{eqn:UL}) and (\ref{eqn:UH}) respectively. The dynamics shows that in both cases the localized initial state loses population to the outer potential wells in an approximately equal manor for both the linear and hysteresis quenches. The strong nonlinear interactions in the initial localized trap repel the condensate symmetrically between the two neighboring traps. Additionally, we notice that in panel (d) the population could be $n_1=n_3\approx 0$, meaning that the final state is exactly equal to the initial state. We have achieved full reversibility with the hysteresis dynamics in these simulations. As shown in panel (c), this is not the case where the populations are always $n_1 \approx n_2 > 0$,  implying that the strong two-body interactions prevent a complete localization of the condensate on a single site.

We now move on to examine the dynamics for the five site system. Without two-body interactions, linearly quenching from $\gamma_i$ to $\gamma_f$ will force the atoms towards the rightmost trap. However from Fig.~\ref{fig:dynamics_as}(e) we see that the occupation is never very much greater than $n_5\approx0.5$, even for the slowest quenching rates considered in the simulation. When the quench rate is fast ($\alpha\sim1$), we find less occupation in both the first and last site, implying the occupation has been spread amongst the remaining sites. In panel (f), the hysteresis counterpart is shown. Now the population should tend towards $n_1\approx1$. However this is not what is found in the numerical simulations. The populations distribute randomly in all sites. In panels (g) and (h), the dynamics is qualitatively different from the $L=3$ scenario. The symmetry between the densities of the two outermost sites is lost completely, and is replaced with a chaotic distributions, largely due to the presence of hyperchaos [see Fig.~\ref{fig:stability2}(d)].

\section{Scaling of Lyapunov exponents with the system size}\label{sec:Size}
In the following we will investigate how the maximal and total number of Lyapunov exponents depend on the system size and initial state, focusing on parameter regimes where the nonlinear interaction can not be neglected, i.e. chaos and hyperchaos are expected in the dynamics. In general Lyapunov exponents depend on the input state of the calculation~\cite{Wolf1985}. Two different initial states, i.e. the groundstate and the localized state, will be examined in detail. 
\begin{figure}
	\centering
	\includegraphics[width=\linewidth]{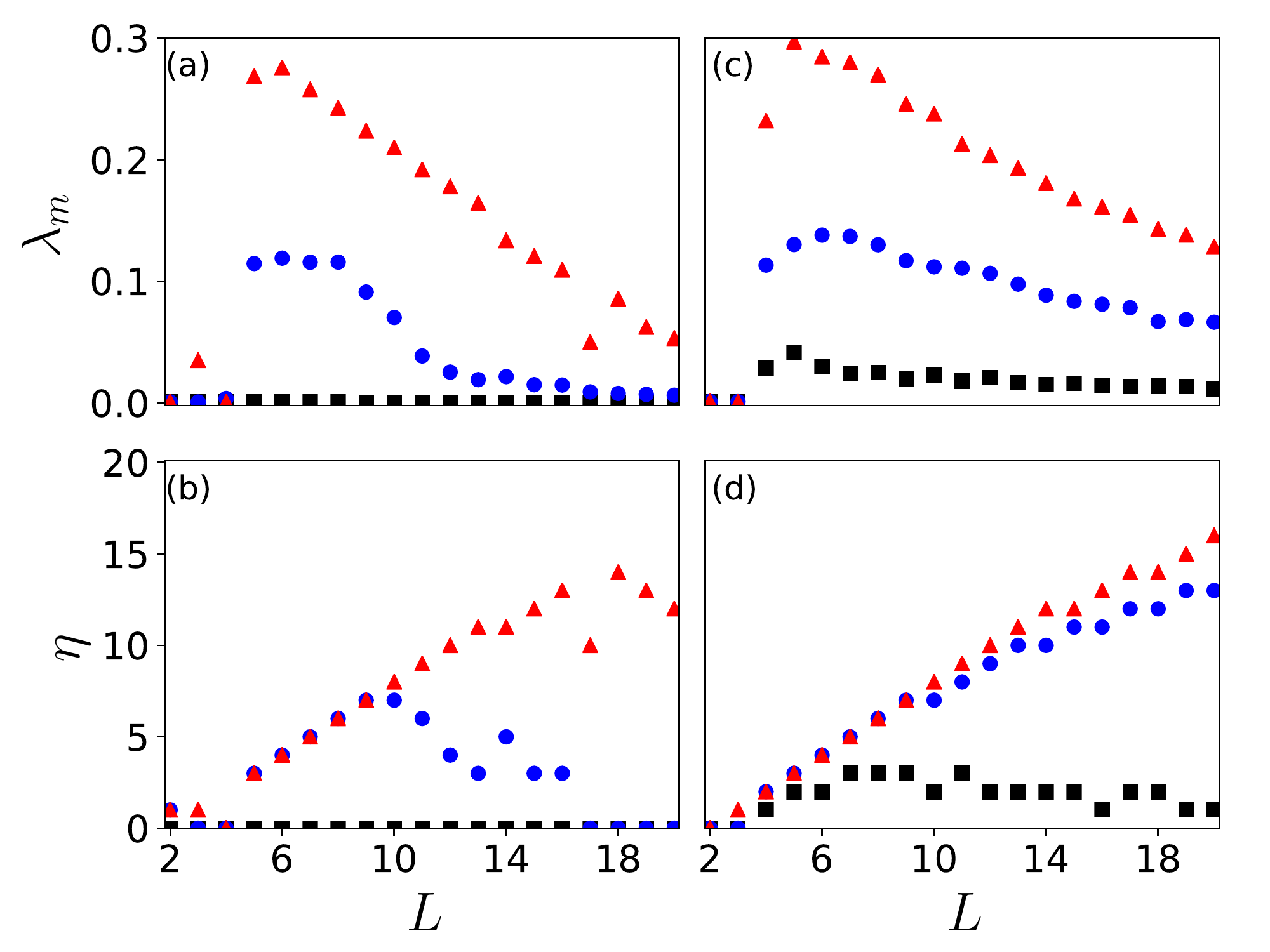}
	\caption{(color online) \textbf{Lyapunov Exponents vs System Size.} The maximum Lyapunov exponent and total number of positive Lyapunov exponents are shown in (a) and (b) for the groundstate configuration. Panels  (c) and (d) show the same quantity for the localized state. The larger $U$ is, the larger the maximal Lyapunov exponent. The maximal Lyapunov exponent decreases with increasing $L$. The number of Lyapunov exponents increases  and then decreases  with increasing $L$.  For the localized state, $\eta$ increases nearly linearly with increasing $L$. In each panel, $U=1$ (square), 3 (circle) and 5 (triangle). }\label{fig:size}
\end{figure}

In Fig. \ref{fig:size}(a) the largest Lyapunov exponent $\lambda_m$ for the groundstate configuration is shown. When $2\le L \le 4$, the values of $\lambda_m$ are small in general. This is due to the fact that chaos has not be triggered [see Fig.~\ref{fig:stability2}(b) and (c)]. When $L>4$, the situation changes as chaos is already found with the given $U$.  We find $\lambda_m$ decreases gradually when $U=3$ and $U=5$ for larger $L$. On the other hand, the total number of positive Lyapunov exponents $\eta$ is seen to increase almost linearly with $L$ when $U=3$ and $U=5$, depicted in Fig.~\ref{fig:size}(b).   Importantly,  $\eta>2$ when $L>4$ for both $U=3$ and $U=5$, i.e. the dynamics is hyperchaotic. On the other hand,  $\eta$ decreases and deviates from the linear dependence on $L$ when $L$ is large, e.g. at $L=10$ when $U=3$ and $L=14$ when $U=5$. In general the linear relation holds up to a larger $L$  for larger $U$. Recently it has been shown that the largest Lyapunov exponents in the BH model can be obtained from the echo dynamics of the condensate~\cite{PhysRevA.96.023624}. Similar technique could be applied to extract the largest Lyapunov exponents studied here.
\begin{figure}
	\centering
	\includegraphics[width=\linewidth]{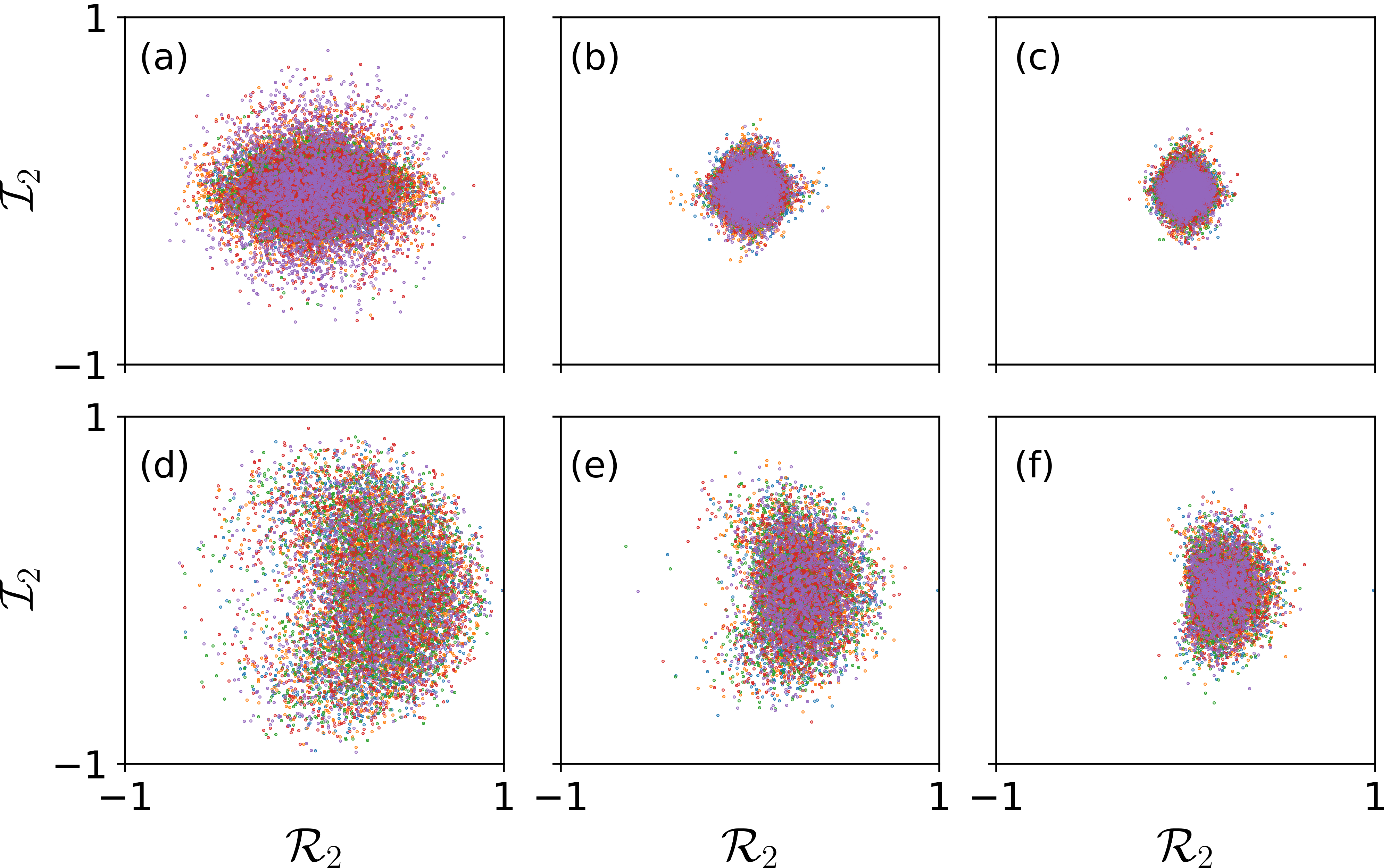}
	\caption{(color online) \textbf{Poincar\'e Sections of the groundstate and localized state on the $\mathcal{U}_1$-plane}. 
		The Poincar\'e sections are shown for the groundstate (a-c) and the localized state  (d-f). Each point represents a numerical realization. We consider $L=5$ (a,d), $L=10$ (b,e), and $L=20$ (c,f). Other parameters are $U=3$ and $\gamma=0$.}
	\label{fig:poincare}
\end{figure}

Figs. \ref{fig:size}(c) and (d) show both $\lambda_m$ and $\eta$ for the the localized state. In this case,  $\lambda_m$ is largest when $L=5$, and decreases with increasing $L$ for $U=3$ and $U=5$. Compared to the groundstate, a visible difference is that $\lambda_m\neq0$ when $U=1$ for the localized state. Their values, however, are smaller than the one for $U=3$ and $U=5$. This implies that it will be difficult to observe chaotic dynamics with this level of nonlinear interactions. On the other hand,  $\eta$ increases with increasing $U$. When $L>10$, $\eta$ still increases with $L$, slightly deviates from the linear scaling with $L$. A similar dependence is also found for stronger nonlinear interactions, as demonstrated with $U=5$ in panel (d). For such state, $\eta>1$ can be seen even with relatively weak interaction (e.g. $U=1$), leading to more pronounced hyperchaotic dynamics. 

The total number of nonlinear differential equations is $2L$ (the real and imaginary parts of $\psi_j$). For conservative systems, the number of positive and negative Lyapunov exponents are the same, and the sum of the Lyapunov exponents is zero. These features can be seen, e.g., in Fig.~\ref{fig:groundstate}(d). Our numerical simulations show that the maximal number of positive Lyapunov exponents is $L-2$ [see Fig.~\ref{fig:size}(b) when $U=5$ and $L\le 11$ and Fig.~\ref{fig:size}(d) when $U=5$ and $L\le 14$.]. As the extended Bose-Hubbard model is a Hamiltonian system, not only the sum of the Lyapunov exponents vanishes, but also conservative quantities, such as the energy and particle number, are found in the dynamics. This indicates that the maximal number of the Lyapunov exponents is $L-2$ but not $L$. For sufficiently large $L$, the total number of positive Lyapunov exponents is smaller than $L-2$, as the nonlinear interaction becomes smaller. For the groundstate, one can estimate the interaction energy for a given site to be $2(U+V)/L^2$ approximately, i.e. the mean local interaction energy decreases with increasing $L$.
\begin{figure}
	\centering
	\includegraphics[width=\linewidth]{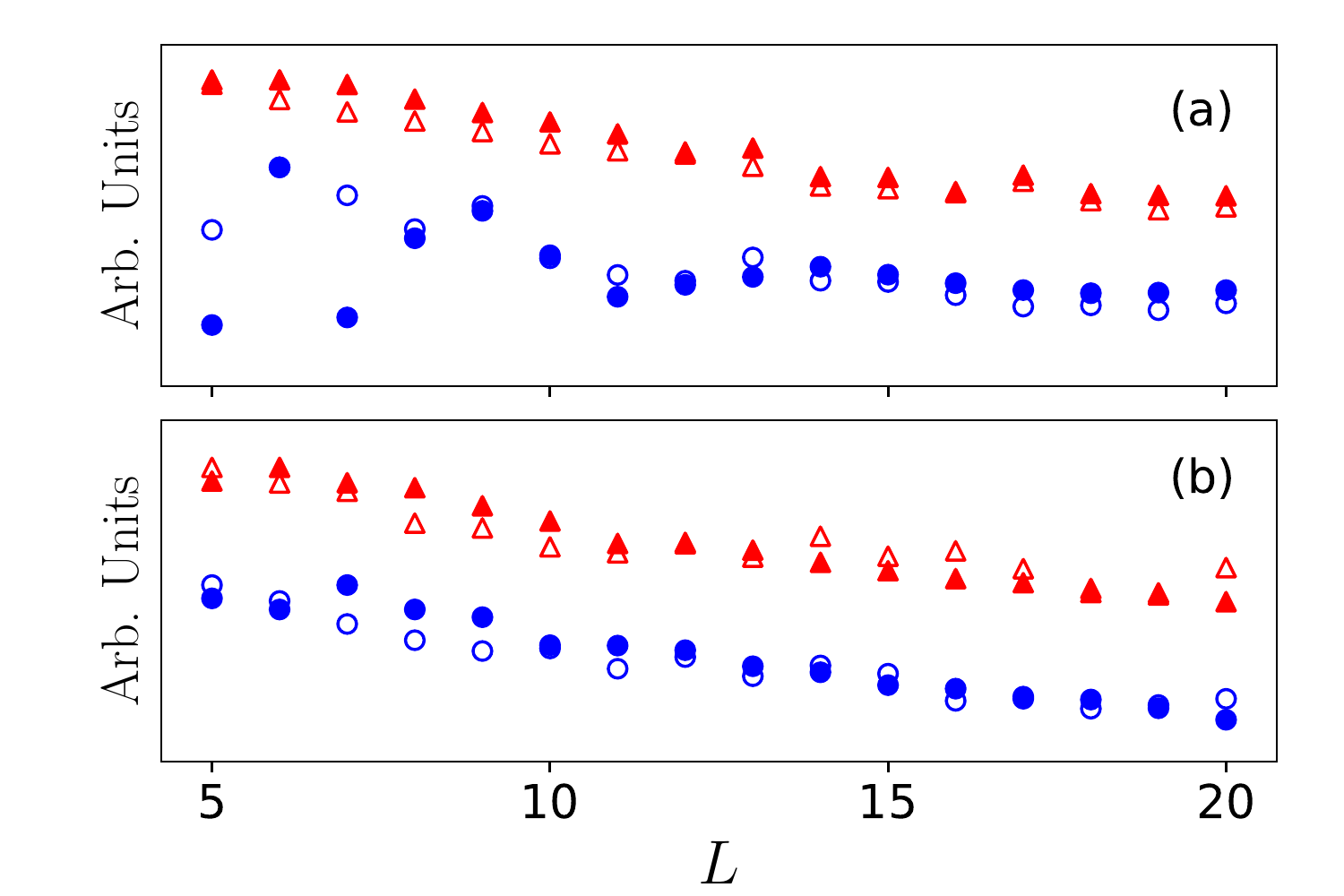}
	\caption{(color online) \textbf{Areas of the Poincar\'e Sections}. We compare the fitted area (open shapes) of the Poincar\'e section with $\lambda_m$ (solid) for both the groundstate (a) and localized state (b), respectively. The blue circles are for $U=3$, and red triangles for $U=5$. In both situations $\gamma=0$. }
	\label{fig:area}
\end{figure}

The chaotic dynamics depends strongly on the largest Lyapunov exponents $\lambda_m$, which is considered as an indication of chaos in the dynamics. To illustrate this, 
the Poincar\'e section on the  $\mathcal{U}_1$  plane for different system sizes is shown in Fig.~\ref{fig:poincare}, showing that profiles of the Poincar\'e section depend on the system size and the initial state. When $L=5$ the area is largest [Fig.~\ref{fig:poincare}(a)] and decrease with increasing $L$ in case of the groundstate [Fig.~\ref{fig:poincare}(b) and (c)]. For different $L$, the profile of the Poincar\'e section is largely symmetric with respect to $\mathcal{R}_2=0$ and $\mathcal{I}_2=0$. In case of the localized state, similar dependence on $L$ is found, as depicted in Fig.~\ref{fig:poincare}(d)-(f). We note two differences compared to the groundstate ones. First, the profile of the Poincar\'e section displays symmetry with respect to $\mathcal{I}_2=0$ but not $\mathcal{R}_2=0$. Second, the areas of the Poincar\'e section in the localized state are slightly larger, as the corresponding $\lambda_m$ is larger [see Fig.~\ref{fig:size}(a) and (c)].
 
The area is largely determined by the largest Lyapunov exponent. To verify this, we find the area of the Poincar\'e section approximately through numerically fitting the Poincar\'e section, shown in Fig.~\ref{fig:area}. For the groundstate, the dependence of the fitted area and $\lambda_m$ on $L$ agrees well when $U=5$. For $U=3$, a good agreement is also found when $L\ge 8$. When $L=5$ and $L=7$, the fitted areas differ largely from the corresponding $\lambda_m$. This discrepancy might be caused by the fact that the relatively weak nonlinear interaction leads to uncertainties in calculating the Lyapunov exponent. For the localized state, the agreement is improved in general for both $U=3$ and $U=5$. This suggests that the discrepancy in the groundstate could be a boundary effect when $L$ is small, as the localized state suffers less from the boundary effect. 

\section{Conclusion and outlook}\label{sec:Conclusion}
We have investigated the chaotic and hyperchaotic dynamics of a one-dimensional Bose-Hubbard chain of Rydberg-dressed BECs in the semiclassical regime.
We have shown that both the groundstate and localized state can have positive Lyapunov exponents, even though the corresponding Bogoliubov spectra are real valued. As a result, small perturbations to these states lead to large fluctuations, which are captured by the quench dynamics. We have found that hyperchaos emerges in both the groundstate and localized states when the nonlinear interaction is strong and $L$ is large. The total number of positive Lyapunov exponents, $\eta$, is bound by $L-2$ ($L\ge 3$). We have shown that $\eta$ grows with the system size $L$ when $U$ is large. So far our investigations are focusing on the semiclassical regime. There has been exploration into the relationships between chaos and quantum entanglement~\cite{Lerose2020a}. Moreover, quantum chaos can be seen by analyzing the statistics of eigenspectra on the Bose-Hubbard model with onsite~\cite{pausch_chaos_2021} and long-range interactions~\cite{kollath_statistical_2010,chen_persistent_2020}. It is therefore worthwhile to explore features of chaos and hyperchaos due to the Rydberg dressed interaction in the quantum regime. 

\section*{Acknowledgements}
The research leading to these results received
funding from EPSRC Grant No. EP/R04340X/1 via the
QuantERA project “ERyQSenS,” UKIERI-UGC Thematic
Partnership No. IND/CONT/G/16-17/73, and the Royal Society
through International Exchanges Cost Share Award No.
IEC$\backslash$NSFC$\backslash$181078. 
R.N. acknowledges DST-SERB for a Swarnajayanti fellowship (File No. SB/SJF/2020-21/19).

\end{document}